# Light extraction from high-*k* modes in a hyperbolic metamaterial

T. GALFSKY[1,2], H. N. S. KRISHNAMOORTHY[1,2], W. NEWMAN[3], E. E. NARIMANOV[4], Z. JACOB[3], V. M. MENON[1,2*]

[1]Department of Physics, City College, City University of New York (CUNY), New York, NY 10031, USA
[2]Department of Physics, Graduate Center, CUNY, New York, NY 10016, USA.
[3]Department of Electrical and Computer Engineering, University of Alberta, Edmonton, AB T6G 2V4, Canada.
[4]Birck Nanotechnology Center, School of Computer and Electrical Engineering, Purdue University, West Lafayette, Indiana 47907, USA.

*Correspondence: Dr. Vinod Menon, Department of Physics, City College, City University of New York (CUNY), New York, NY 10031, USA
E-mail: vmenon@ccny.cuny.edu

**ABSTRACT**

Hyperbolic Metamaterials (HMMs) have recently garnered much attention because they possess the ability for broadband manipulation of the photon density of states and sub-wavelength light confinement. These exceptional properties arise due to the excitation of electromagnetic states with high momentum (high-k modes). However, a major hindrance to practical applications of HMMs is the difficulty in coupling light out of these modes because they become evanescent at the surface of the metamaterial. Here we report the first demonstration of out-coupling of high-k modes in an active HMM using a high-index contrast bulls-eye grating. Quantum dots embedded inside the metamaterial are used for local excitation of high-k modes. This demonstration of light out-coupling from quantum dots embedded in a HMM could pave the way for developing practical photonic devices using these systems.

## INTRODUCTION

Artificially engineered sub-structured materials having hyperbolic dispersion are known as hyperbolic metamaterials (HMMs) or indefinite media. Amongst the various metamaterial systems being explored, HMMs have received most attention lately due to the simplicity of the structures, and a host of unique applications that span the frequency range from the microwave down to the UV while having non-magnetic permittivity ($\mu=1$) [1–4]. HMMs composed of metal-dielectric stacks support multiple surface plasmons polaritons (SPPs) which are responsible for their unique properties [5,6]. Unlike other plasmonic metamaterials the plasmonic response of HMMs is broadband and can be tuned away from the resonant frequency of the individual building blocks. Quantum emitters such as dyes and quantum dots (QDs) placed inside or on top of HMMs have been shown to exhibit enhanced decay rates and giant Purcell factor enhancement [7–10]. However most of the emission from sources in the near field of these materials is radiated predominantly into the structure which is prohibitive for many applications in the far-field unless an out-coupling mechanism is applied [11,12]. It has been previously shown that dipole emitters placed inside a multilayered structure experience a much stronger Purcell factor enhancement than on top of the structure [9], however placing the emitters inside the structure also results in very weak detectable emission. A possible solution to this was proposed theoretically using a bulls-eye grating [12]. In this work we show for the first time experimental verification of out-coupling of high-$k$ modes generated locally by dipole emitters embedded inside a HMM using a high refractive index contrast bulls-eye grating.

**MATERIALS AND METHODS**

The structure (shown schematically in Figure 1a) consists of alternating layers of aluminum-oxide ($Al_2O_3$) at ~20nm thickness and silver (Ag) at ~12nm with ultra-thin layers of Germanium (Ge) (~1-2nm thick) in between the two acting as a wetting layer for silver in order to achieve optical quality smooth films [13,14] (see SI section 1). The unit cell of $Al_2O_3$/Ge/Ag layers is defined as one period. At the first stage of the fabrication, four periods (4P) were deposited on a glass cover-slip by electron-beam evaporation and terminated with a thin (6$nm$) $Al_2O_3$ spacer. CdSe/ZnS quantum dots (QDs) were then spin-coated on top of the spacer layer. The deposition of layers continued with another 6$nm$ $Al_2O_3$ spacer and three more periods completing the seven periods (7P) structure. The HMM was terminated with a thin capping layer of $Al_2O_3$ (~3nm) in order to protect the last Ag layer from oxidation. We determined the exact thickness of the dielectric and metallic layers from cross-sectional imaging with a transmission electron microscope (TEM) (Figure 1b) and the optical constants of thin layers of Ag/Ge and $Al_2O_3$ were determined from spectroscopic ellipsometery measurements. Since the permittivity of each layer was known from ellipsometry and the exact fill fraction of metal in the structure was known from TEM imaging, we calculated the components of the effective anisotropic permittivity using effective medium theory (Figure 1c and SI section 2). We see that starting at a wavelength of 420nm the parallel ($\parallel$) and perpendicular ($\perp$) permittivity components have opposite signs which lends a hyperboloid shape to the iso-frequency contour according to the dispersion relation equation

$$\frac{k_x^2 + k_y^2}{\varepsilon_\perp} + \frac{k_z^2}{\varepsilon_\parallel} = \frac{\omega^2}{c^2} \quad (1)$$

The shape of the hyperbolic iso-frequency contour is schematically drawn in the overlay of Figure 1c along with the spherical iso-frequency contour of air, depicting the issue of mode mismatch between free space and HMM modes. The spectrum of the QDs used in the experiment is shown in Figure 1c as a dashed line.

The number of high-$k$ modes supported in a metal-dielectric stack is determined by the number of confined dielectric layers in the structure [5]. These coupled plasmonic modes create enhancement in the local photonic density of states and provide a radiative decay channel for quantum emitters placed on top or inside the HMM [7,12,15]. Fluorescence lifetime imaging microscopy (FLIM) is used to test the effect of high-$k$ modes on the spontaneous emission rate of the QDs. The lifetime of the 7P and 4P structures was compared to a control sample consisting of a single period (1P), and QDs on glass substrate. Figure 1d shows the lifetime distribution for the four samples. The measured reduction in average lifetimes is comparable to previous reports [8,9], with the 7P structure showing decay rate enhancement factor of 8.8 in comparison to glass. In addition to the experimental measurement, the expected decay rate enhancement was analytically calculated from the local photonic density of states of the exact structure showing a good match to the experimentally obtained values (see SI section 3).

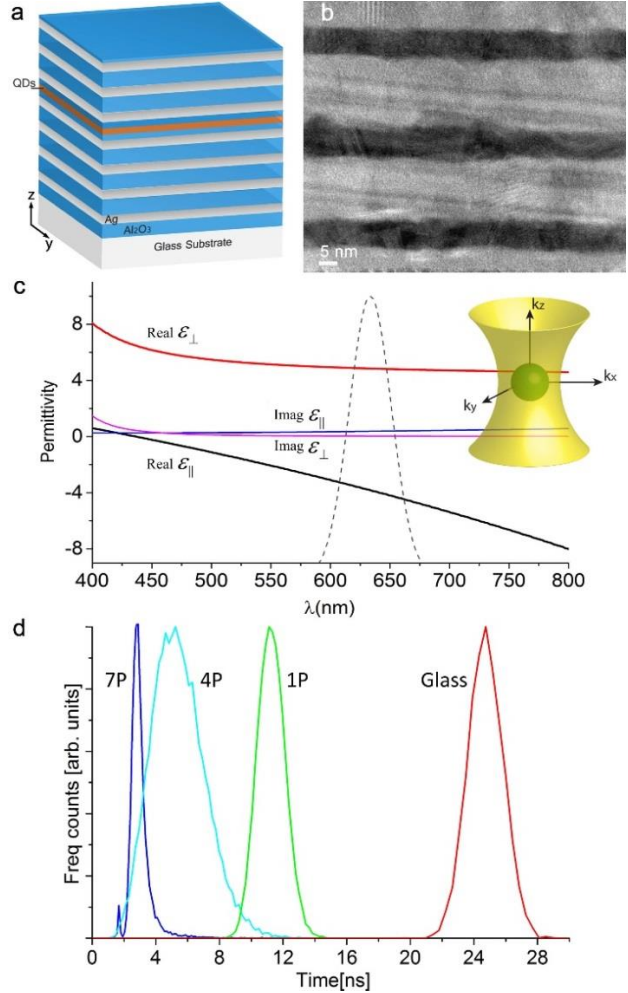

Figure 1. (a) Schematic of the multilayered structure composed of Ag/Al$_2$O$_3$ (b) Cross-sectional Transmission Electron Microscope (TEM) image showing smooth continuous films of Ag (dark) and Al$_2$O$_3$ (bright). (c) Permittivity of the structure as calculated from Effective Medium Theory based on ellipsometrically determined dielectric constants and metal fill fraction determined from cross-sectional TEM. The dotted line corresponds to the emission spectrum of the QDs. Inset: The hyperbolic iso-frequency surface of an ideal HMM overlaid on the spherical iso-frequency surface of free space. (d) Lifetime distribution of the QDs on glass, control sample - 1 period (1P), four period HMM (4P), and inside a seven period HMM structure (7P).

Having obtained the exact thickness and the optical constants of the layers also allowed us to simulate the anticipated optical properties of the fabricated structure using finite-element-method (FEM) software. Figure 2a shows the simulated electric field generated by a vertical electric dipole inside the 7P HMM, showing emission inside the structure having a typical X-shape known as resonance cone [12,16] which is a unique property of hyperbolic media. The hyperbolic shape of the iso-frequency contour in the material also affects the Poynting vector, which is normal to the iso-frequency contour, and lies within the resonance cone. Thus, the energy flow due to high-$k$ modes is inherently directed towards the surface of the HMM [12]. However, the electric field becomes evanescent at the top interface resulting in nearly no detectable emission in the far-field. The dipole emission is internally reflected at the top and bottom interfaces as can be seen in Figure 2a. To translate the evanescent field into propagating waves a grating structure is required. Due to the difficulty in extending the Green's function approach to arbitrary geometries and random dipole orientations a numerical approach was used for the grating design. Multi-parameter FEM simulations for grating height and periodicity for several compatible materials were optimized to maximize out-coupling from the HMM within an angle range of ± 45 degrees with respect to the normal (Figure 2b). Despite having losses in the visible range Ge was selected over compatible lossless dielectrics because of its high index of refraction which allows it to act as a high contrast grating (HCG) with air [17] thus aiding the out-coupling of high-*k* modes (see SI section 4). From Figure 2b one also sees that the out-coupled power is more sensitive to the grating period than to the grating height. The bulls-eye shape of the grating was selected in order to provide radial symmetry to the out-coupled beam [18]. The emission pattern from the 7P HMM with an optimally designed bulls-eye grating (Δ=125*nm*, *h*=60*nm*) is shown in Figure 2c. The far-field emission pattern of the optimized grating structure for a vertically oriented dipole and horizontally oriented dipole is shown in Figure 2d. It is clear that the vertically oriented dipole achieves much better coupling to the HMM high-*k* modes due to its stronger TM component [10,16]. The spectral bandwidth of the optimized grating is ~50*nm* which covers emission spectrum of the QDs used in the present experiment (see SI section 4).

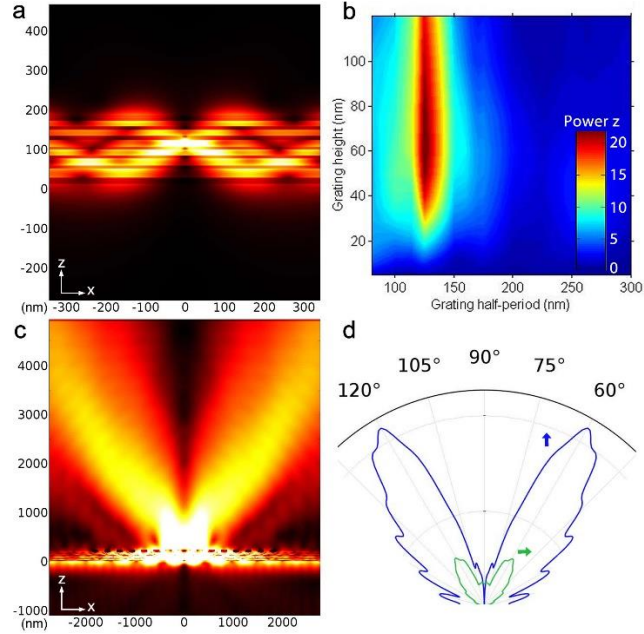

Figure 2. (a) Simulation of the electric field generated by a vertical dipole placed inside the 7P HMM. Black arrow represents the vertical dipole. (b) Out-coupled power in an angle range of -45 to 45 degrees as a function of grating period and height. (c) Out-coupling of the evanescent field by an optimized Ge grating with half-period 125nm and height of 60nm (d) Far-field emission (arb. units) of vertical (blue) and horizontal (green) dipoles.

## RESULTS AND DISCUSSION

To demonstrate the effect of grating period on out-coupling efficiency, bulls-eye gratings with half-periods Δ=125, 150, 170, 200, 250, 265 and 300*nm* were patterned by electron-beam lithography onto a 100nm thick PMMA resist layer which was spin-coated on top the HMM. The gratings defined on PMMA were used as a mask for depositing Ge and realizing the high-contrast bulls-eye gratings (see SI section 5). Scanning electron microscope (SEM) images of the Δ=150nm grating and an array of gratings with different periodicities are shown in Figure 3.

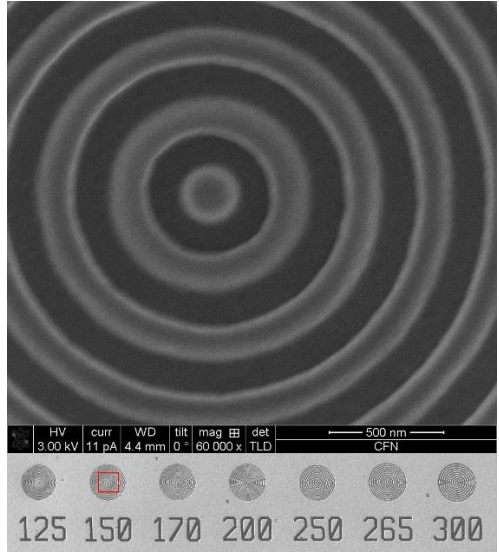

Figure 3. Scanning electron microscope (SEM) image of Δ=150nm bulls-eye grating (top); array of gratings with different periods (bottom).

Photoluminescence (PL) measurements were carried out using an inverted confocal microscope setup coupled to an avalanche photo diode detector. The QDs were pumped at 440nm by a solid state pulsed laser and the emission spectrum was separated by a dichroic filter (see SI section 5). The gratings were imaged with a 60X air objective with NA of 0.85.

Figure 4a shows the PL image of the gratings with different periods. It is clearly seen that only the region where the grating is present out-couples the light from the HMM to the far field. The dark background elucidates the evanescent nature of the high-$k$ modes which do not propagate to the far-field. The bright spots in emission seen inside the gratings are due to the presence of clusters of QDs formed during spin-coating and were not included in the analysis of the PL intensity images.

The ratio of the average PL intensity from the different grating periods to the un-patterned areas is presented in Figure 4b together with numerical calculation of this ratio from FEM simulations. The simulations use the experimentally measured PL spectrum of the QDs and take into account the random orientation of the dipoles. In both simulation and experiment the amount of out-coupled light increases for smaller gratings with the Δ=125*nm* grating yielding the highest output. The sharp emission contrast observed between the grating region and the un-patterned background in the HMM (Figure 4a) is due to out-coupling of multiple high-$k$ modes. In comparison, the 1P structure shows far smaller contrast between the background and the grating region due to the excitation of only one SPP mode.

Another feature of the emission which is observed most clearly in the Δ=125nm grating is the dark circle at the center. In order to explain this feature we turn our attention to the simulation of the electric field shown in Figure 4c. In this simulation of a vertical dipole placed below the center of the grating it can be seen that the light emitted follows a cone-like emission pattern (the horizontal dipole emits the same way with lesser out-coupling efficiency). If a straight line is traced from the final far-field direction it appears as if the beam originated from a spot with a radius of ~1μm around the grating center which approximately corresponds to the radius of the dark center in the image (Figure 4c inset).

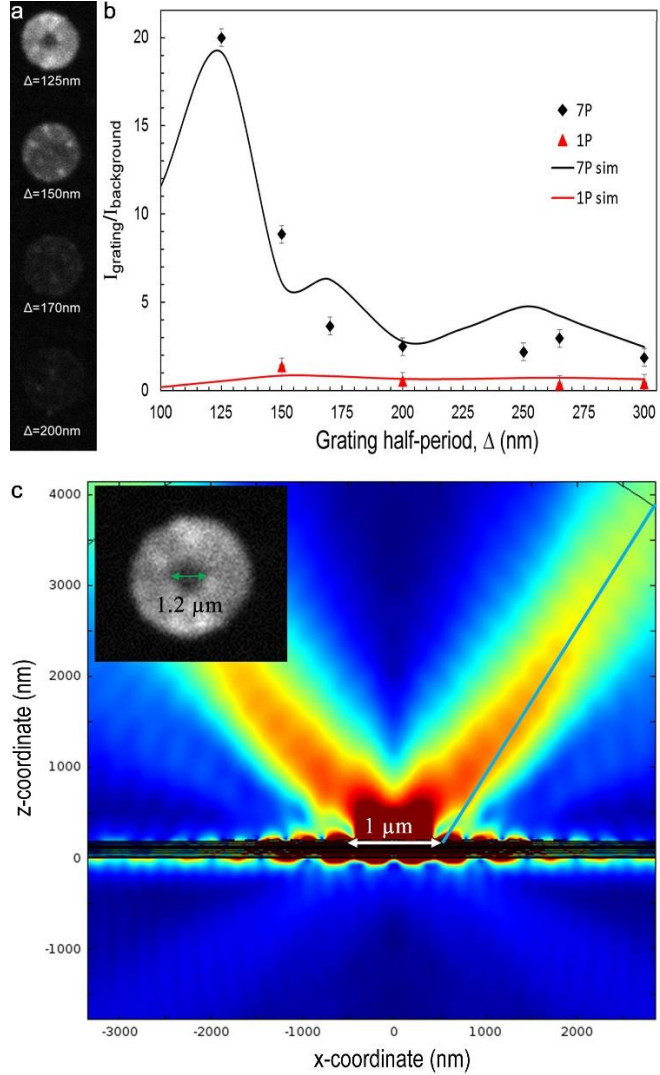

Figure 4. (a) Confocal microscope image of PL emission from the bulls-eye gratings with half period, $\Delta$=125,150,170,200nm on the 7P HMM. The measurements were carried out over a scan area of $10\mu m \times 40\mu m$. Details of the measurement technique can be found in the SM section 5. (b) Experimentally measured ratio between the intensity at the center of the grating to that of non-patterned background for the 7P and the 1P structures for different grating periods. Also shown in solid lines are the calculated ratios from simulations for the two cases using ensemble approximation. (c) FEM simulation of dipole emission from a $\Delta$=125nm grating. The beam curves above the center of the grating. When light is collected in the far-field the origin of the beam appears to be shifted. The inset shows the confocal image of the emission pattern having a dark center.

## CONCLUSIONS

We present a conclusive demonstration of light extraction from high-$k$ modes in a HMM embedded with QDs. A high refractive index contrast bulls-eye grating structure fabricated on top of the active HMM was used to enhance the light extraction from the embedded emitters by a factor of 20 over the un-patterned area. This large enhancement in light-extraction in conjunction with the reduction in spontaneous emission lifetime is indicative of high-k modes supported by the HMM which are being out-coupled to the far-field. We note that even with optimized grating design, the total number of out-coupled photons is eventually restricted by the material loss in metal-dielectric structures such as the one shown here. In this context, the recent work on low-loss HMMs with extremely high PDOS [19] provides an ideal platform to implement such out-coupling mechanisms for embedded emitters needed for quantum nano-photonics applications. The approach presented in our work for grating design and execution is general and applicable for HMMs of different materials and emitters at various frequency ranges. Even for very broadband emitters such as nitrogen vacancy centers in diamond, one can easily extend the present approach using chirped gratings. High refractive index contrast gratings can be realized with a variety of CMOS compliant materials and promises to be even more efficient in the near infrared where the loss in materials such as Ge and Si is much lower. Control of spontaneous emission and extraction of light into the far-field from active HMM is an important step to achieving practical photonic devices such as sub-wavelength lasers, superluminescent LEDs and single photon sources.

## ACKNOWLEDGMENTS

Research carried out in part at the Center for Functional Nanomaterials, Brookhaven National Laboratory, which is supported by the U.S. Department of Energy, Office of Basic Energy Sciences, under Contract No. DE-AC02-98CH10886. The authors acknowledge support form Army Research Office (ARO) (W911NF1310001); Helmholtz Alberta Initiative; National Science and Engineering Research Council (NSERC) Canada; National Science Foundation (NSF) (DMR 1120923); ARO Multidisciplinary Research Initiative (MURI).

The authors thank Mircea Cotlet and Huidong Zang for assistance with imaging and Yun Yu for assistance with AFM measurements.

# Light extraction from high-*k* modes in a hyperbolic metamaterial: Supplementary Information

T. GALFSKY[1,2], H. N. S. KRISHNAMOORTHY[1,2], W. NEWMAN[3], E. E. NARIMANOV[4], Z. JACOB[3], V. M. MENON[1,2*]

[1]Department of Physics, City College, City University of New York (CUNY), New York, NY 10031, USA
[2]Department of Physics, Graduate Center, CUNY, New York, NY 10016, USA.
[3]Department of Electrical and Computer Engineering, University of Alberta, Edmonton, AB T6G 2V4, Canada.
[4]Birck Nanotechnology Center, School of Computer and Electrical Engineering, Purdue University, West Lafayette, Indiana 47907, USA.

*Correspondence: Dr. Vinod Menon, Department of Physics, City College, City University of New York (CUNY), New York, NY 10031, USA
E-mail: vmenon@ccny.cuny.edu

## 1. QUALITY OF LAYERS

The optical quality of the layers is highly important in order to obtain bulk plasmon-polariton modes and avoid local scattering. A silver layer with thickness of less than 20*nm* tends to form percolated films when deposited on dielectric materials. It has been shown that a seed layer of Ge acts as an efficient wetting layer for Ag and results in an optically smooth surface with very low roughness [1,2]. We used a Ge seed layer (<2*nm*) which allows fabrication of ultra-smooth thin silver films without significantly altering the overall optical properties of the HMM. Fig. S1a shows an Atomic Force Microscope (AFM) image of the top surface of the unit cell in our structure: 10*nm* Ag layer deposited on top of the Ge seed layer on 15nm of $Al_2O_3$ with a glass substrate. The top layer Ag film is found to have surface roughness factor $R_{RMS} = 0.1nm$ with maximum peak to valley roughness of 0.5*nm*. For a single $Al_2O_3$ layer of similar thickness a surface roughness factor of $R_{RMS} = 0.3nm$ was measured with maximum peak to valley roughness of <2*nm*. These values were corroborated by Transmission Electron Microscope (TEM) images. Fig. S1b shows a cross-sectional TEM image of the 7P structure showing an $Al_2O_3$ layer sandwiched between two metallic layers. The individual layers are well defined with minimal variation over distance scales up to 10*μm* in the direction parallel to the layers. All layers were deposited by electron-beam evaporation in vacuum pressure $< 1 \cdot 10^{-5}$ *Torr* and at a deposition rate of $0.3 Å/sec$.

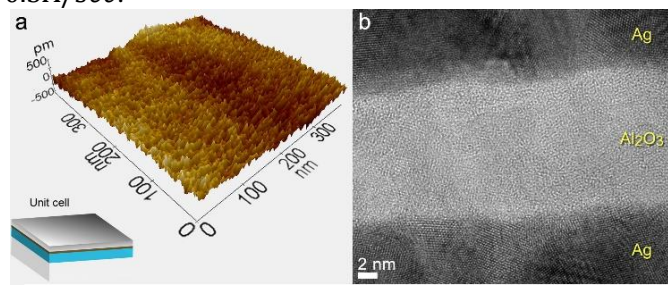

Figure S1. (a) AFM image of the top surface of a unit cell consisting of Ag/Ge/$Al_2O_3$ (top to bottom) on a glass substrate. (b) TEM image of an $Al_2O_3$ layer sandwiched between two Ag layers. The Ge wetting layer is indistinguishable in the TEM image.

## 2. EFFECTIVE MEDIUM THEORY

A metal-dielectric stack can be validly approximated by a homogenous anisotropic medium when the unit cell thickness is smaller than $\lambda/10$ and if it consists of at least 4 periods [3]. In such case the parallel ( $\parallel$ ) and perpendicular ( $\perp$ ) permittivity components of the medium with respect to the layers are given by

$$\varepsilon_\parallel = \rho \varepsilon_m + (1-\rho)\varepsilon_d, \qquad (S1)$$

$$\varepsilon_\perp = \frac{\varepsilon_m \varepsilon_d}{\rho \varepsilon_d + (1-\rho)\varepsilon_m}, \qquad (S2)$$

where $\varepsilon_m$ and $\varepsilon_d$ are the complex permittivity components of the metal and dielectric respectively and $\rho$ is the fill fraction of the metal, $\rho = t_m/(t_m + t_d)$. In our 4P and 7P structures $\rho \simeq 0.36$ which places the transition point into the hyperbolic regime around wavelength $\lambda = 420 nm$.

## 3. LOCAL PHOTONIC DENSITY OF STATES AND RADIATIVE ENHANCEMENT

Although in an ideal structure approximated by EMT, hyperbolic dispersion corresponds to infinite local photonic density of states (LPDOS), in a realistic structure that includes finite layers and losses, the LPDOS is discretized into finite number of plasmonic modes. We use a dyadic Green's function approach [3–5] to calculate the wavelength resolved LPDOS as a function of the normalized wave-vector parallel to the layers ($k_x/k_0$) for our four period (4P) and seven period (7P) structures with an emitter positioned on top of the 4P and at the center of the 5$^{th}$ dielectric layer of the 7P. In the weak coupling regime a quantum emitter can be treated as a radiating point dipole. Since our system contains randomly oriented dipoles we must take into account the contributions of dipole orientations parallel ($\parallel$) and perpendicular ($\perp$) to the layers. Following the work of Ford & Weber [5] we can write the LPDOS for both orientations as following:

$$\rho_\parallel = \frac{1}{k_z} \frac{k_\parallel}{\sqrt{\varepsilon_d} k_0} \left[ 1 + r_s e^{2ik_z d} + (1-r_p e^{2ik_z d}) \frac{k_z^2}{\varepsilon_d k_0^2} \right], \qquad (S3)$$

$$\rho_\perp = \frac{1}{k_z} \left( \frac{k_\parallel}{\sqrt{\varepsilon_d} k_0} \right)^3 \left( 1 + r_p e^{2ik_z d} \right), \qquad (S4)$$

Where $k_z = \sqrt{k_\parallel^2 - \varepsilon_d k_0^2}$, $k_\parallel = \sqrt{k_x^2 + k_y^2}$ is the wave-vector parallel to the layers (x-y plane), $k_0$ is the wave-vector in free-space, $\varepsilon_d$ is the permittivity of the layer in which the dipole is embedded, $r_s$ and $r_p$ are the reflection coefficients for TE and TM polarized light calculated by transfer matrix method. In the ensemble approximation we can write the total LPDOS as

$$\rho(\lambda, k_\parallel) = \frac{1}{3}\rho_\perp + \frac{2}{3}\rho_\parallel, \qquad (S5)$$

$$\frac{\Gamma}{\Gamma_0} = (1-\eta) + \eta \operatorname{Re}\left[ \int_0^\infty \rho(\lambda, k_\parallel) dk_\parallel \right], \qquad (S6)$$

Where $\eta$ is the quantum yield of the emitter ($\eta \sim 1$.), and $\Gamma_0$ is the radiative decay rate of the dipole in air. Fig. S2a and S2b show the calculated LPDOS for the dipoles in the 4P and 7P structures where high-$k$ modes are distinguished bright bands with normalized wave-vectors $k_x/k_0 > 2$. These modes provide a radiative decay channel for the dipoles. Fig. S2c and S2d show the calculated radiative decay rate enhancement $(\Gamma/\Gamma_0 \equiv \beta)$ as a function of wavelength. In the wavelength range of our QDs the average enhancement for the 4P structure is calculated to be $\beta_{4P} = 4.8$ and for the 7P structure $\beta_{7P} = 8.75$ which well corresponds with the experimentally obtained values of $\beta_{4P,\exp} = 4.84$ and $\beta_{7P,\exp} = 8.8$.

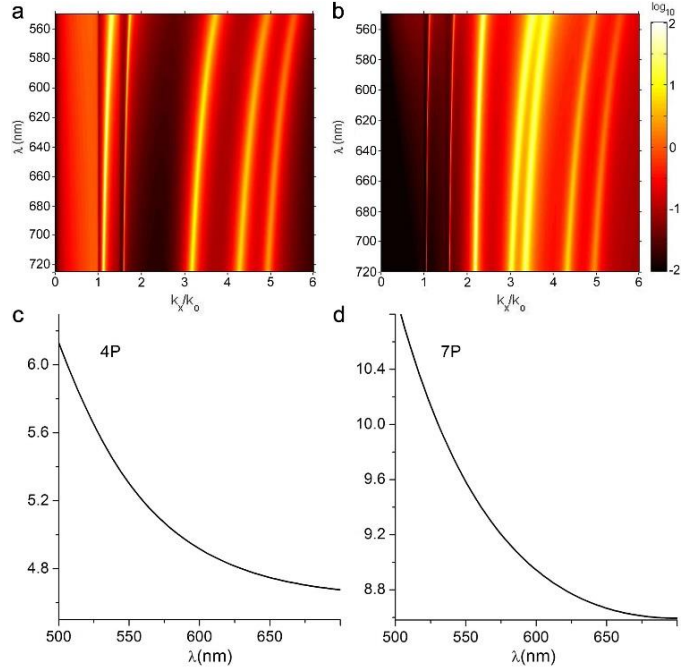

Figure S2. Top Panel: Local Photonic Density of States as a function of wavelength and wave-vector parallel to the layers (a) 4P structure (b) 7P structure. Bottom Panel: Radiative decay enhancement as a function of wavelength calculated for (c) 4P and (d) 7P.

## 4. GRATING DESIGN

FEM numerical simulations to determine the out-coupling efficiency were performed using commercial software COMSOL Multiphysics™. In the simulations, a double-parameter sweep of the grating half-period and height are performed to determine the maximum power in the upwards ($\hat{z}$) direction within an emission half-angle of $45^0$. The grating optimization simulations are done for a dipole located directly under the grating, emitting at the center wavelength of our QDs, $\lambda_{em} = 635nm$. We take into account the isotropic distribution of dipole moments of the QDs by solving for both perpendicular ($\perp$) and horizontal ($\parallel$) dipole orientations with respect to the layers and averaging the out-coupled power according to the ensemble approximation

$$P_{ens} = (P_{\parallel} + 2P_{\perp})/3, \qquad (S7)$$

We examined the out-coupling efficiency of several available materials for the grating such as PMMA, $Al_2O_3$, Ag, and Ge. Fig. S3 shows a comparison the out-coupling parameter space for an $Al_2O_3$ grating (Fig. S3a) and a Ge grating (Fig. S3b). The maximum out-coupled power from the Ge grating at height of 60$nm$ and half-period 125nm is more than double the maximum out-coupled power from the $Al_2O_3$ grating at height of 120$nm$ and half-period 200$nm$. Due to its large real part of the refraction index (~4) Ge acts as a high contrast grating (HCG) assisting in translating the high-$k$ modes into propagating waves in the far-field.

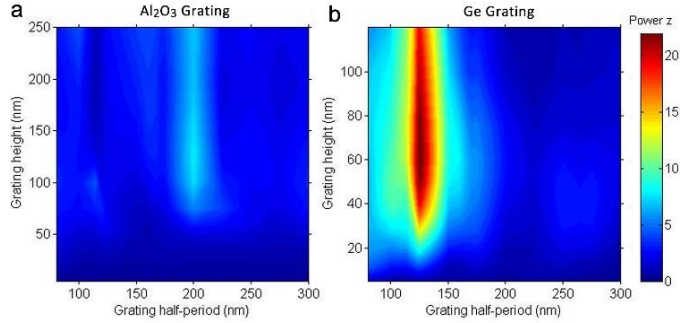

Figure S3. Out-coupled power in the z direction with a $45^0$ half-angle. (a) $Al_2O_3$ grating. (b) Ge grating. The Ge grating provides more than double of the maximum efficiency of the $Al_2O_3$ grating.

Fig. S4 plots the full wavelength response for the maximum output parameters of a Ge grating, $\Delta=125nm$ and $60nm$ grating height. The out-coupled power reaching the far-field within half-angle of $45^0$ is normalized by the total power emitted by a dipole in air with a flat frequency spectrum. It is seen that even with the poor output of horizontally oriented dipoles it is possible to achieve directional output that exceeds the total power of an emitter in free space. The output peak has a FWHM of $50nm$ which is suitable for out-coupling of emitters such as QDs with a similar FWHM spectral width of $\sim 40nm$. For emitters with a broader spectrum, such as nitrogen vacancy centers in diamond, a chirped grating design can be applied.

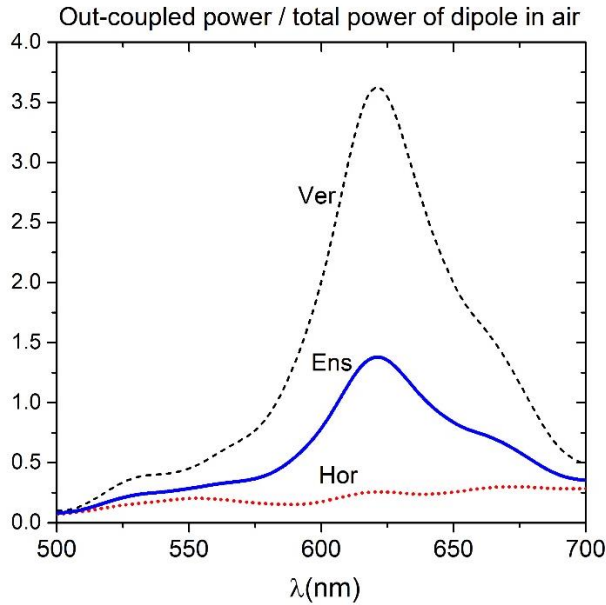

Figure S4. Grating wavelength response. Ver, Hor, and Ens denote the response for a vertical dipole, horizontal dipole, and an ensemble.

## 5. FABRICATION AND IMAGING

**Fabrication.** $Al_2O_3$, Ge, and Ag layers were deposited on pre-cleaned glass cover slips by e-beam evaporation using a Lekser PVD 75 Thin Film System while keeping the pressure $< 1 \cdot 10^{-5}$ *Torr*. After growing four periods of $Al_2O_3$/Ge/Ag terminating with a $6nm$ thick $Al_2O_3$ spacer. CdSe/ZnS core/shell quantum dots in Toluene purchased from NN Labs were diluted to a concentration of 25% by volume and spin-coated on top of the structure forming a layer of thickness $\sim 25 \pm 5nm$ then covered by another $6nm$ of $Al_2O_3$ and the remaining three periods, again terminating with a thin ($\sim 3nm$) $Al_2O_3$ capping layer. At this point PMMA A2 was spin coated on the top layer at 7000RPM and patterned using JEOL JBX6300-FS e-beam lithography system. The pattern was developed in 3:1 MIBK and Isopropanol for 60 seconds. Ge was deposited on top and the PMMA film was removed by acetone lifted-off process leaving behind the inverse grating in Ge. Although the optimal grating design called for $60nm$ high grating, our fabrication resulted in a thinner grating of 30±10*nm* as determined by DekTak 150 stylus surface profiler Although non-optimal, this grating height is still within the maximum out-coupling design of the Δ=125nm grating (see Fig. S3b) since the amount of out-coupled power is relatively insensitive to the grating height.

**Imaging.** Fluorescence lifetime imaging microscopy measurements were performed on an inverted confocal microscope (Oxford X71) coupled to a diode pumped solid state laser delivering 440*nm*, 90*p*sec pulses, at 8MHz repetition rate. The sample's luminescence was spectrally separated from the laser by a Semrock RazorEdge 532 long pass filter (LPF) and detected by an MPD Picoquant Avalanche Photodiode (APD) coupled to a PicoHarp 300 time analyzer. See Fig. S4 for a schematic of the setup. The laser spot size on the sample is $0.5\mu m \times 0.5\mu m \times 2\mu m$ and the pinhole placed before the APD has a diameter of $75\mu m$.

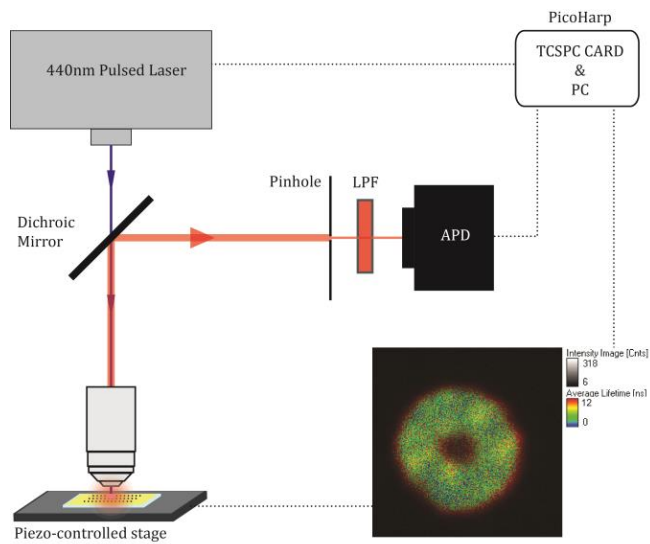

Figure S5. Schematic of fluorescence lifetime imaging setup.